\documentstyle[amssymb,prl,aps,epsfig]{revtex}

\begin{document}

\title{Probing inhomogeneities in type II superconductors by means of thermal
fluctuations, magnetic fields and isotope effects}
\author{T. Schneider}
\address{Physik-Institut der Universit\"{a}t Z\"{u}rich, Winterthurerstrasse 190,\\
CH-8057 Z\"{u}rich, Switzerland}
\maketitle

\begin{abstract}
Type II superconductors, consisting of superconducting domains
embedded in a normal or insulating matrix, undergo a rounded phase
transition. Indeed, the correlation length cannot grow beyond the
spatial extent of the domains. Accordingly, the thermodynamic
properties will exhibit a finite size effect. It is shown that the
specific heat and penetration depth data of a variety of type II
superconductors, including cuprates, exhibit the characteristic
properties of a finite size effect, arising from domains with
nanoscale extent. The finite size scaling analysis reveals
essential features of the mechanism. Transition temperature and
superfluidity increase with reduced domain size. The combined
finite size and isotope effects uncover the relevance of local
lattice distortions.
\end{abstract}
\bigskip


\bigskip
To appear in J. Superc., Proceedings Dynamic Inhomogeneities in
Complex Systems, June 2003, Bled

\section{Introduction}
Since the discovery of superconductivity in cuprates by Bednorz
and M\"{u}ller\cite{bed} a tremendous amount of work has been
devoted to their characterization. The issue of inhomogeneities
and their characterization is essential for several reasons,
including: First, if inhomogeneity is an intrinsic property, a
re-interpretation of experiments, measuring an average of the
electronic properties, is unavoidable. Second, inhomogeneity may
point to a microscopic phase separation, i.e. superconducting
grains, embedded in a non-superconducting matrix. Third, there is
neutron spectroscopic evidence for nanoscale cluster formation and
percolative superconductivity in various
cuprates\cite{mesot,furrer}. Fourth, nanoscale spatial variations
in the electronic characteristics have been observed in underdoped
Bi$_{2}$Sr$_{2}$CaCu$_{2}$O$_{8+\delta }$ with scanning tunnelling
microscopy (STM)\cite{liu,chang,cren,lang}. They reveal a spatial
segregation of the electronic structure into 3nm diameter
superconducting domains in an electronically distinct background.
On the contrary, a large degree of homogeneity has been observed
by Renner and Fischer\cite{renner}. As STM is a surface probe the
relevance of these observations for bulk and thermodynamic
properties remains to be clarified. Fifth, in
YBa$_{2}$Cu$_{3}$O$_{7-\delta }$, MgB$_{2}$, 2H-NbSe$_{2}$ and
Nb$_{77}$Zr$_{23}$ considerably larger grains have been
established. The magnetic field induced finite size effect
revealed lower bounds ranging from $L=182$ to $814$\AA\
\cite{tsfs}. Sixth, since the change of the lattice parameters
upon oxygen isotope exchange is negligibly small, the occurrence
of a significant change in the inhomogeneities spatial extent,
will provide clear evidence for the relevance of local lattice
distortions in mediating superconductivity\cite {tsrkhk}.

Here we review the attempts to probe the granularity of type II
superconductors by means of thermal fluctuations and magnetic
fields\cite {tsfs,tsrkhk}. It is well-known that systems of finite
extent, i.e. isolated superconducting grains, undergo a rounded
and smooth phase transition\cite {fisher}. As in an infinite and
homogeneous system the transition temperature $T_{c}$ is
approached the correlation length $\xi $ increases strongly and
diverges at $T_{c}$. However, when superconductivity is restricted
to grains with length scale $L_{i}$ in direction $i$, $\xi _{i}$
cannot grow beyond $L_{i}$. In type II superconductors, exposed to
a magnetic field $H_{i}$, there is an additional limiting length
scale $L_{H_{i}}=\sqrt{\Phi _{0}/\left( aH_{i}\right) }$ with
$a\approx 3.12$, related to the average distance between vortex
lines \cite{tsfs}. Indeed, as the magnetic field increases, the
density of vortex lines becomes greater, but this cannot continue
indefinitely, the limit is roughly set on the proximity of vortex
lines by the overlapping of their cores. Due to these limiting
length scales the phase transition is rounded and occurs smoothly.
Indeed, approaching $T_{c}$ from below the transverse correlation
length $\xi _{i}^{t}$ in direction $i$ and its real space
counterpart $\xi _{i}^{-}=\sqrt{\xi _{j}^{t}\xi _{k}^{t}}$, where
$i\neq j\neq k$, increase strongly. However, due to the limiting
length scales $L_{i}$ and $L_{H_{i}}=\sqrt{\Phi _{0}/\left(
aH_{i}\right) }$, it is bounded and cannot grow beyond
\begin{equation}
\left.
\begin{array}{c}
\xi _{i}^{-}\left( t_{p}\right) =\xi _{0i}^{-}\left| t_{p}\right|
^{-\nu
}=L_{i}, \\
\sqrt{\xi _{i}^{-}\left( t_{p}\right) \xi _{j}^{-}\left(
t_{p}\right) }=\sqrt{\xi _{0i}^{-}\xi _{0j}^{-}}\left|
t_{p}\right| ^{-\nu }=\sqrt{\Phi _{0}/\left( aH_{k}\right)
}=L_{H_{k}},\ i\neq j\neq k,
\end{array}
\right\}  \label{eq1}
\end{equation}
where $t_{p}=1-T_{p}/T_{c}$ and $\nu $ denotes the critical
exponent of the correlation lengths. Beyond the mean-field
approximation it differs from $\nu =1/2$ and $a\approx 3.12$ is a
universal constant\cite{tsfs}. In this context it is important to
recognize that the confinement effect of the magnetic field in
direction $i$ on fluctuations within a region $L_{H_{i}}$ acts
only in the plane perpendicular to the field. As a remnant of the
singularity at $T_{c}$ the thermodynamic quantities exhibit a so
called finite size effect, i.e. a maximum or an inflection point
at $T_{p} $. Two limiting regimes, characterized by
\begin{equation}
L_{H_{i}}=\sqrt{\frac{\Phi _{0}}{aH_{i}}}=\left\{
\begin{array}{c}
<\sqrt{L_{j}L_{k}} \\
>\sqrt{L_{j}L_{k}}
\end{array}
,i\neq j\neq k\right\} ,  \label{eq2}
\end{equation}
can be distinguished. For $L_{H_{i}}<\sqrt{L_{j}L_{k}}$ the
magnetic field induced finite size effect limits $\sqrt{\xi
_{j}^{-}\xi _{k}^{-}}$ to grow beyond $L_{H_{i}}$, while for
$L_{H_{i}}>\sqrt{L_{j}L_{k}}$ the superconducting grains set the
limiting length scales. Since $L_{H_{i}}$ can be tuned by the
strength of the applied magnetic field, both limits are
experimentally accessible. $L_{H_{i}}<\sqrt{L_{j}L_{k}}$ is
satisfied for sufficiently high and $L_{H_{i}}>\sqrt{L_{j}L_{k}}$
for low magnetic fields. Thus, the occurrence of a magnetic field
induced finite size effect requires that the magnetic field and
the length scales of the superconducting grains satisfy the lower
bound $H_{i}>\Phi _{0}/\left( aL_{j}L_{k}\right) $. Since
superconductors fall in the experimentally accessible critical
regime into the 3D-XY universality class\cite{book}, with known
critical exponents and critical amplitude combinations, we take
these properties for granted\cite {peliasetto}. They include the
exponents
\begin{equation}
\alpha =2-3\nu =-0.013,\ \nu =0.671  \label{eq3}
\end{equation}
and the critical amplitude combinations
\begin{equation}
k_{B}T_{c}=\frac{\Phi _{0}^{2}}{16\pi ^{3}}\frac{\xi
_{0k}^{-}}{\lambda _{0i}\lambda _{0j}},\ \ i\neq j\neq k,\ \
A^{\pm }V_{c}^{-}=\left( R^{\pm }\right) ^{3},V_{c}^{-}=\xi
_{0i}\xi _{0j}\xi _{0k}  \label{eq4}
\end{equation}
where
\begin{equation}
\frac{A^{+}}{A^{-}}=1.07,\ R^{-}=0.815  \label{eq5}
\end{equation}
$\alpha $ and $A^{\pm }$ are the critical amplitude of the
specific heat singularity, $c=\left( A^{\pm }/\alpha \right)
\left| t\right| ^{-\alpha }+B^{\pm }$, while $\xi _{0i}^{-}$ and
$\lambda _{0i}$ are the critical amplitude of the correlation
length and penetration depth in direction $i$. In the homogeneous
system these length scales diverge as $\xi _{i}^{-}=\xi
_{0i}^{-}\left| t\right| ^{-\nu }$, $\lambda _{i}=\lambda
_{0i}\left| t\right| ^{-\nu /2}$ and $V_{c}^{-}$ is the
correlation volume.
\section{Specific heat}

\begin{figure}[tbp]
\centering
\includegraphics[totalheight=6cm]{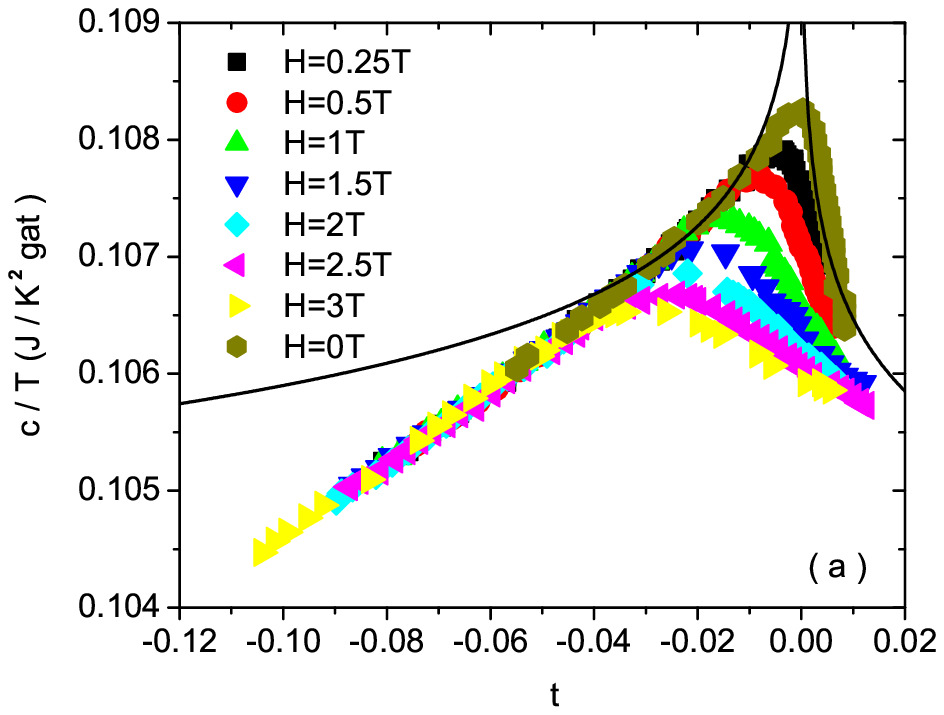}
\includegraphics[totalheight=6cm]{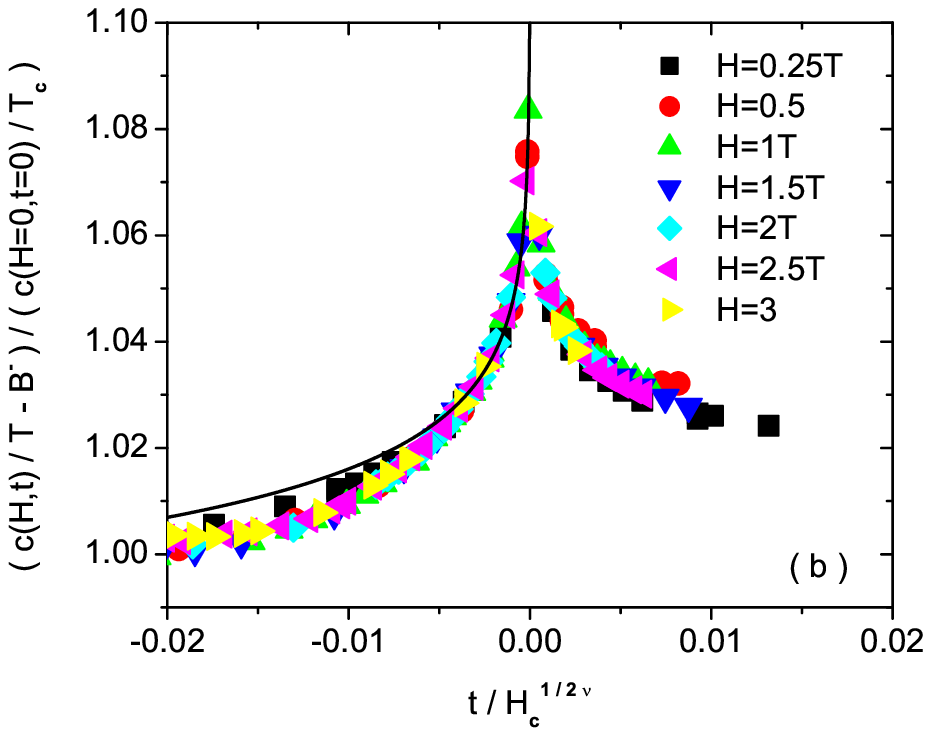}
\caption{(a): \ Specific heat coefficient $c/T$ versus $t$ of
YBa$_{2}$Cu$_{3}$O$_{6.9}$ with $T_{c}=92.56$ K at various
magnetic fields derived from the data of Roulin {\em et
al.}\protect\cite{roulin}. The solid lines are
$c/T=\widetilde{A}^{\pm }\left| t\right| ^{-\alpha
}+\widetilde{B}^{\pm }$ with $\alpha $ and $A^{+}/A^{-}$ given by
Eqs.(\ref{eq3}) and the parameters listed in the text. The
deviations of the zero field data from these lines reveal around
$t=0$ the finite size effect due to inhomogeneities; (b):\ $\left(
c\left( T,H_{c}\right) /T-\widetilde{B}^{-}\right) /\left(
\widetilde{A}^{-}\left| t\right| ^{\alpha }\right) $ versus
$t/H_{c}^{1/2\nu }$ for YBa$_{2}$Cu$_{3}$O$_{6.9}$ with
$T_{c}=92.56$ K derived from the data of Roulin {\em et al.}
\protect\cite{roulin} with $H$ in T. The solid line is $\left(
c\left( T,H_{c}\right) /T-\widetilde{B}^{-}\right) /\left(
\widetilde{A}^{-}\left| t\right| ^{-\alpha }\right) =0.957\left(
-t/H_{c}^{1/2}\right) ^{\alpha }$ with $\alpha =-0.013$,
confirming the divergence of the scaling function in the limit
$\pm y\rightarrow 0$.} \label{fig1}
\end{figure}

In Fig.\ref{fig1}a we displayed the data of Roulin {\em et
al.}\cite{roulin} for the temperature dependence of the heat
coefficient of a YBa$_{2}$Cu$_{3}$O$_{6.9}$ single crystal with
$T_{c}=92.56$K at various magnetic fields applied parallel to the
c-axis ($H_{c}$). As a remnant of the zero field singularity,
there is for fixed field strength a broad peak adopting its
maximum at $T_{p}$ which is located below $T_{c}$. As $T_{p}$
approaches $T_{c}$, the peak becomes sharper with decreasing $H$
and evolves smoothly to the zero-field singularity, smeared by the
inhomogeneity induced finite size effect. Since $T_{p}$ decreases
systematically with increasing field, condition (\ref{eq2})
applies in the form $L_{H_{c}}<\sqrt{L_{a}L_{b}}$and there is a
magnetic field induced finite size effect. As the data of Roulin
{\em et al.}\cite{roulin} are rather dense and extend close to
zero field criticality, a detailed finite size scaling analysis
appears to be feasible. In Fig.\ref{fig1}b we plotted the data in
terms of $\left( c\left( T,H_{c}\right) /T-B^{-}\right) /\left(
A^{-}\left| t\right| ^{-\alpha }/\alpha \right) $ versus
$t/H_{c}^{1/2\nu }$. Apparently the data falls on a single curve,
which is the finite size scaling function $g\left( y\right) $
defined in terms of
\begin{equation}
\left( \frac{c\left( T\right) }{T}-\widetilde{B}^{-}\right)
/\left( \widetilde{A}^{-}\left| t\right| ^{-\alpha }\right)
=g\left( y\right) ,\ y=t\left( \Phi _{0}/\left( aH_{c}\left( \xi
_{0ab}^{-}\right) ^{2}\right) \right) ^{1/2\nu }.  \label{eq6}
\end{equation}
A finite size effect requires that $g(-1)=1$ at $T_{p}$, where
$y=y_{p}=-1$, while $g\left( \pm y\rightarrow 0\right) \propto
\left| y\right| ^{-\left| \alpha \right| }$ holds for $\alpha <0$
and $g\left( \pm y\rightarrow \infty \right) =1$. To qualify the
remarkable collapse of the data we note that the estimates for
$\widetilde{A}^{-}=A^{-}/\alpha $ and $B^{-}$ have been derived
from the zero field data displayed in Fig.\ref{fig1}a.\ The solid
line for $t<0$ is $c/T=\widetilde{A}^{-}\left| t\right| ^{-\alpha
}+\widetilde{B}^{-}$with $\widetilde{A}^{-}=A^{-}/T_{c}\alpha
=-0.073$(J/K$^{2} $gat)$,\ \
\widetilde{B}^{-}=0.179$(J/K$^{2}$gat),
$\widetilde{B}^{+}=0.181$(J/K$^{2}$gat), $T_{c}=92.56$K and the
critical exponent $\alpha $ $=-0.013$, to indicate the
inhomogeneity induced deviations from the leading zero field
critical behavior of \ perfect YBa$_{2}$Cu$_{3}$O$_{6.9}$. Since
the data collapse onto the finite scaling function has been
achieved without any arbitrary fitting parameter, the existence of
the magnetic field induced finite size effect is well confirmed.
Furthermore, Fig.\ref{fig1}a shows how the broad anomaly in the
specific heat coefficient sharpens, while the maximum height at
$T_{p}\left( H_{c}\right) $ increases with reduced field strength,
evolving smoothly to the zero field peak, rounded by
inhomogeneities. From the finite size scaling plot displayed in
Fig.\ref {fig1}b we also deduce that with $\xi _{0ab}^{-}=14.4$A,
that at $y_{p}$, where $\left| t\right| /H_{c}^{1/2}=-0.013$, the
scaling function is close to $1$, as required. Furthermore, the
solid line which is $\left( c\left( T,H_{c}\right)
/T-\widetilde{B}^{-}\right) /\left( \widetilde{A}^{-}\left|
t\right| ^{-\alpha }\right) =0.957\left( -t/H_{c}^{1/2}\right)
^{\alpha }$, confirms the divergence of the scaling function in
the limit $\pm y\rightarrow 0$.
The shift of $T_{p}$ with increasing $H_{c}$ is according to
Eq.(\ref{eq1}) given by
\begin{equation}
\left| t_{p_{i}}\right| =\left( \frac{a\xi _{0i}^{-}\xi
_{0j}^{-}H_{k}}{\Phi _{0}}\right) ^{1/2\nu }.  \label{eq7}
\end{equation}
\begin{figure}[tbp]
\centering
\includegraphics[totalheight=6cm]{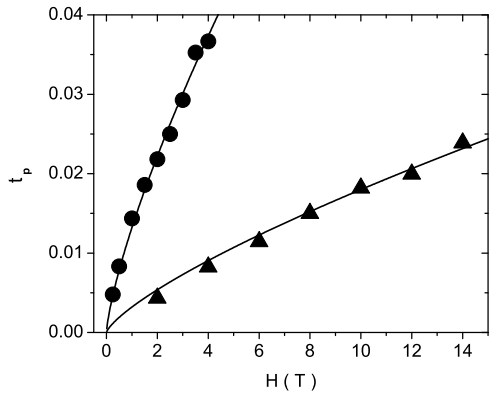}
\caption{$\left| t_{p}\right| $ versus $H_{c}$ ($\bullet $) and
$H_{ab}$ ($\blacktriangle $) for YBa$_{2}$Cu$_{3}$O$_{6.9}$ with
$T_{c}=92.56$K derived from the data of Roulin {\em et al.}
\protect\cite{roulin}. The solid lines are $\left| t_{p}\left(
H_{ab}\right) \right| =0.0033\ H_{ab}^{1/2\nu }$ and $\left|
t_{p}\left( H_{c}\right) \right| =0.0132\ H_{c}^{1/2\nu }$ with
$\nu $ listed in Eq.(\ref{eq3}).} \label{fig2}
\end{figure}
In Fig.\ref{fig2} we show t$_{p}$ versus $H_{ab}$ and $H_{c}$ for
YBa$_{2}$Cu$_{3}$O$_{6.9}$ single crystal with $T_{c}=92.56$K
derived from the data of Roulin {\em et al.}\cite{roulin}. The
solid lines are $\left| t_{p}\left( H_{ab}\right) \right| =0.0033\
H_{ab}^{1/2\nu }$ and $\left| t_{p}\left( H_{c}\right) \right|
=0.0132\ H_{c}^{1/2\nu }$ with $H$, revealing that the critical
regime is attained. With $\nu $ given by Eq.(\ref{eq3}) and
$a=3.12$ these fits yield for the critical amplitudes the
estimates
\begin{equation}
\sqrt{\xi _{0ab}^{-}\xi _{0c}^{-}}=5.7\text{\text{\AA }},\ \xi
_{0ab}^{-}=14.4\text{\text{\AA }, }V_{c}^{-}=\left( \xi
_{0ab}^{-}\right) ^{2}\xi _{0c}^{-}=469\text{\text{\AA }}^{3}.
\label{eq8}
\end{equation}
To estimate the correlation volume from the specific heat
coefficient in terms of the universal relation, $\ A^{\pm
}V_{c}^{-}=\left( R^{\pm }\right) ^{3}$(Eq.(\ref{eq4}) we note
that $A^{-}=T_{c}\alpha $ $\widetilde{A}^{-}$,
$A^{-}$(cm$^{-3}$)$=\left( 10^{7}/k_{B}/V_{gat}\right)
A^{-}$(mJ/K/cm$^{3}$), $V_{gat}=8$cm$^{3}$ and
$\widetilde{A}^{-}=-0.073$(J/K$^{2}$gat), corresponding to the
solid line in Fig.\ref{fig1}a, give $A^{-}=7.96\ 10^{-4} $\AA
$^{-3}$, yielding with $R^{-}=0.815$ (Eq.(\ref{eq5})) the
correlation volume $V_{c}^{-}\approx 680$\AA $^{3}$, which is
reasonably close to the value derived from the magnetic field
induced finite size effect (Eq.(\ref {eq8})). Since at the lowest
attained fields $H_{c}=0.25$T and $H_{ab}=2$T a shift from $T_{c}$
to $T_{p}$ is still present (see Fig.\ref{fig1}a), we derive from
Eq.(\ref{eq2}) for the length scales of the sample inhomogeneities
the lower bounds
\begin{equation}
L_{ab}>515\text{\text{\AA }},\ \
\sqrt{L_{ab}L_{c}}>182\text{\text{\AA }}. \label{eq9}
\end{equation}
As in zero field the correlation volume $V_{c}^{-}\left|
t_{p}\right| ^{-3\nu }$ cannot grow beyond the volume $V_{i}$ of
the inhomogeneities, this limiting volume scale is obtained from
$V_{i}=V_{c}^{-}\left| t_{p}\right| ^{-3\nu }$ and $t_{p}$
evaluated in zero field. With $V_{c}^{-}=469$\AA $^{3}$ and
$\left| t_{p}\right| =0.0025$ taken from Fig.\ref{fig1}a, we
obtain the estimates $V_{i}=7.5\ 10^{7}$\AA$^{3}$ and
$V_{i}^{1/3}=422$\AA\ which is consistent with the aforementioned
lower bounds.

\begin{figure}[tbp]
\centering
\includegraphics[totalheight=6cm]{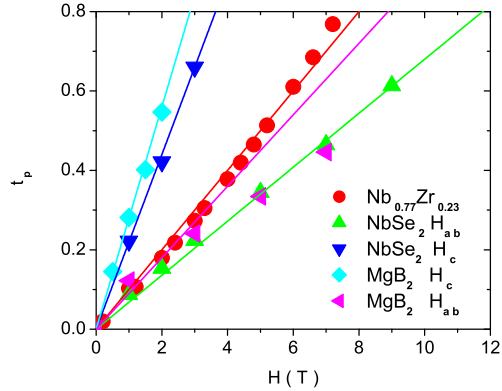}
\caption{$\left| t_{p}\right| $ versus $H$ for
Nb$_{77}$Zr$_{23}$\protect\cite {mirmelstein},
2H-NbSe$_{2}$\protect\cite{sanchez} and
MgB$_{2}$\protect\cite{lyard} derived from the respective
references. The straight lines are Eq.(\ref{eq10}) with the
parameters listed in Table I.} \label{fig3}
\end{figure}

Next we turn to Nb$_{77}$Zr$_{23}$\cite{mirmelstein},
2H-NbSe$_{2}$\cite {sanchez} and MgB$_{2}$\cite{lyard}, type II
superconductors supposed to have comparative large correlation
volumes. In Fig.\ref{fig3} we displayed $\left| t_{p}\right| $
versus $H$ derived from the respective experimental data. In
contrast to the corresponding plot for nearly optimally
YBa$_{2}$Cu$_{3}$O$_{7-\delta }$ (Fig.\ref{fig2}), the data points
to a linear relationship. Consequently, the critical regime, where
$\left| t_{p}\right| $ $\propto $ $H^{1/2\nu \text{ }}$with $\nu
\approx 2/3$ holds (Eq.(\ref{eq7})), is not attained.
Nevertheless, there is clear evidence for a magnetic field induced
finite size effect, because $T_{p}$ shifts monotonically to a
lower value with increasing magnetic field. Since the data points
to an effective critical exponent $\nu \approx 1/2$ which applies
over an unexpectedly extended range, we use Eq.(\ref{eq7}) with
$\nu =1/2$ to derive estimates for the amplitude of the respective
transverse correlation lengths and the correlation volume in terms
of
\begin{equation}
\left| t_{p}\right| =b_{i}H_{i}=\frac{\xi _{0j}^{-}\xi
_{0k}^{-}}{L_{H_{i}}^{2}}=\frac{aH_{i}\xi _{0j}^{-}\xi
_{0k}^{-}}{\Phi _{0}}, \label{eq10}
\end{equation}
with $a=3.12$.
The respective straight lines in Fig.\ref {fig3} are this relation
with the parameters listed in Table I. For comparison we included
the corresponding parameters for
YBa$_{2}$Cu$_{3}$O$_{6.9}$\cite{roulin},
YBa$_{2}$Cu$_{3}$O$_{6.6}$, (K,Ba)BiO$_{3}$\cite{KBa} and
HgBa$_{2}$Ca$_{2}$Cu$_{3}$O$_{7-\delta }$\cite{carrington}where
the critical regime is attained. It is evident that
Nb$_{77}$Zr$_{23}$, 2H-NbSe$_{2}$ and MgB$_{2}$ are type II
superconductors with comparatively large correlation lengths.
Compared to YBa$_{2}$Cu$_{3}$O$_{6.9}$ and
YBa$_{2}$Cu$_{3}$O$_{6.6}$ the correlation volume is 3 orders of
magnitude larger, which renders the amplitude of the specific heat
singularity very weak (see Eq.(\ref{eq4}). Nevertheless, the
unambiguous evidence for the magnetic field induced finite size
effect reveals that fluctuations, even though not critical, are at
work. For this reason there is no critical line $T_{c2}\left(
H\right) $ of continuous phase transitions, but a line
$T_{p}\left( H\right) $ where the specific heat peak, broadening
and decreasing with increasing field, adopts its maximum value and
the correlation length attains the limiting magnetic length scale
$L_{H_{i}}$. Because the fluctuations are also subject to the
finite size effect arising from inhomogeneities with length scale
$L_{j}$, the magnetic finite size effect is observable as long as
$\sqrt{L_{j}L_{k}}>L_{H_{i}}$. The resulting lower bounds for the
length scale of inhomogeneities affecting the thermodynamic
properties are also included in Table I. Noting that these bounds
stem from studies where no attempt was made to explore the low
field behavior, it is conceivable that the actual length scale of
the inhomogeneities is much larger. To our best knowledge, the
only absolute reference stems from the finite size scaling
analysis of the zero field specific heat data of nearly optimally
doped YBa$_{2}$Cu$_{3}$O$_{7-\delta }$, where $L$ was found to
range from $290$ to $419$\AA \cite{book,housten}. Interestingly
enough, the largest bound found here applies to the cubic
superconducting alloy Nb$_{77}$Zr$_{23}$.
\bigskip

\begin{tabular}{|c|c|c|c|c|c|c|c|}
\hline & T$_{c}$(K) & $\xi _{0}^{-},\sqrt{\xi _{0ab}^{-}\xi
_{0c}^{-}}$(\AA) & $\xi _{0ab}^{-}$(\AA) & $\gamma $ &
V$_{c}^{-}$(A$^{3}$) & L,L$_{ab}$(\AA) & $\sqrt{L_{ab}L_{c}}$(\AA) \\
\hline Nb$_{77}Zr_{23}$ & 10.8 & 55 & - & 1 & 1.7 10$^{5}$ & $>$
814 & - \\ \hline
2H-NbSe$_{2}$ & 7.1 & 43 & 94 & 4.8 & 1.7 10$^{5}$ & $>$ 258 & $>$ 258 \\
\hline MgB$_{2}$ & 35 & 52 & 110 & 4.5 & 2.9 10$^{5}$ & $>$ 364 &
$>$ 258 \\ \hline (K,Ba)BiO$_{3}$ & 31.6 & 50 & - & 1 & 1.25
10$^{5}$ & $>$ 258 & - \\ \hline YBa$_{2}$Cu$_{3}$O$_{6.95}$ &
92.6 & 5.7 & 14.4 & 6.4 & 4.7 10$^{2}$ & $>$ 515 & $>$ 182 \\
\hline YBa$_{2}$Cu$_{3}$O$_{6.6}$ & 64.2 & (7.3) & 32.6 & (20) &
1.7 10$^{3}$ & $>$ 576 & - \\ \hline
HgBa$_{2}$Ca$_{2}$Cu$_{3}$O$_{7-\delta }$ & 111.1 & - & 12 & (52)
& - & $>$ 346 & - \\ \hline
Bi$_{2.12}$Sr$_{1.71}$Ca$_{1.22}$Cu$_{1.95}$O$_{8+\delta }$ & 85.7
& - & - & - & - & $<$ 69 & $<$ 69 \\ \hline
\end{tabular}

\bigskip

Table I:Summary of the estimates for the critical amplitudes of
the correlation lengths, correlation volume $V_{c}^{-}=\left( \xi
_{0ab}^{-}\right) ^{2}\xi _{0c}^{-}$, anisotropy $\gamma $ $=\xi
_{0ab}^{-}/\xi _{0c}^{-}$, lower bounds for the length scales\
$L_{ab}$ and $\sqrt{L_{ab}L_{c}}$ of inhomogeneities, derived from
experimental data by means of the magnetic field induced finite
size effect. For YBa$_{2}$Cu$_{3}$O$_{6.6}$ we used $\gamma =20$,
taken from Janossy {\em et al.}\cite{janossy} to estimate $\xi
_{0ab}^{-}$.

\begin{figure}[tbp]
\centering
\includegraphics[totalheight=6cm]{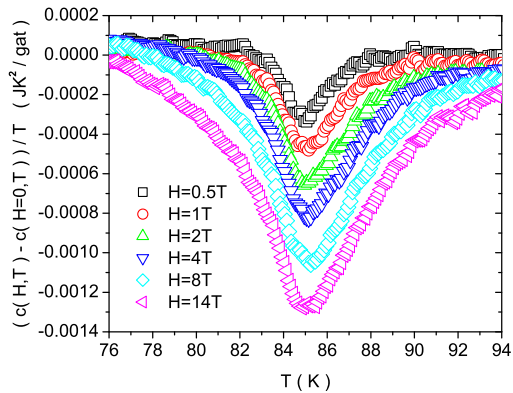}
\caption{$\left( c\left( H_{c},T\right) -c\left( H_{c}=0,T\right)
\right) /T$ \ versus $T$ \ of a
Bi$_{2.12}$Sr$_{1.71}$Ca$_{1.22}$Cu$_{1.95}$O$_{8+\delta }$ single
crystal taken from Junod {\em et al.} \protect\cite{junodbi2212}.
The magnetic field is parallel to the c-axis} \label{fig4}
\end{figure}

\bigskip
On the contrary, the specific heat measurements reveal that in the
Bismuth- and Thallium based cuprates $T_{p}$ appears not to shift
up to $14$T, for fields applied parallel or perpendicular to the
c-axis \cite {junod,mirmelsteinx,junodbi2212}. To derive detailed
estimates we consider the single crystal data for the specific
heat coefficient of Junod {\em et al}.\cite{junodbi2212} for
Bi$_{2.12}$Sr$_{1.71}$Ca$_{1.22}$Cu$_{1.95}$O$_{8+\delta }$. To
reduce artifacts of the huge background we displayed in
Fig.\ref{fig4} the data in terms of $\left( c\left( H,T\right)
-c\left( H=0,t\right) \right) /T$ versus $T$ for fields applied
parallel to the c-axis. Apparently, there is no shift of the peak
around $T_{p}$ $=85.1$K with increasing magnetic field up to
$H=14$T. Noting that the same behavior was found for fields
perpendicular to the c-axis\cite{junodbi2212}, these data lead
with inequality (\ref{eq2}) to the upper bounds
\begin{equation}
L_{ab}<69\text{\AA, }\sqrt{L_{ab}L_{c}}<69\text{\AA},
\label{eq10b}
\end{equation}
uncovering nanoscale superconducting grains, consistent with the
length scale of the inhomogeneities observed with STM
spectroscopy\cite {liu,chang,cren,lang}. As STM is a surface
probe, our analysis establishes that these grains are not merely
an artefact of the surface, but a bulk property with spatial
extent, giving rise to a rounded thermodynamic phase transition
which occurs smoothly.

\section{Penetration depth}

\begin{figure}[tbp]
\centering
\includegraphics[totalheight=6cm]{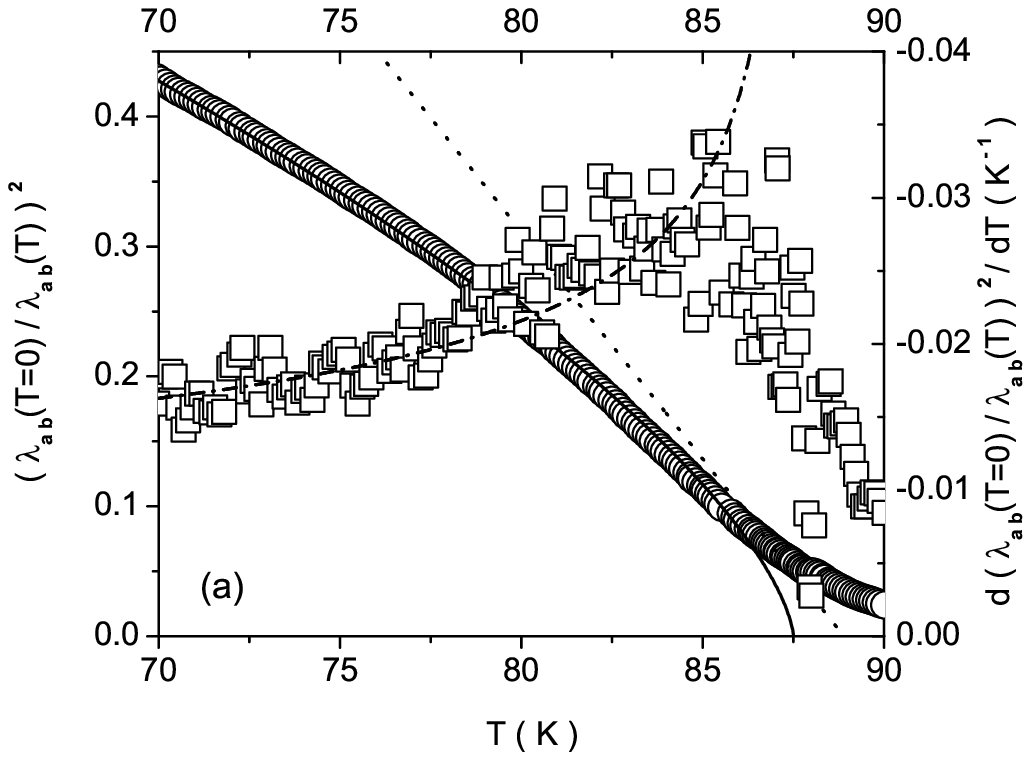}
\includegraphics[totalheight=6cm]{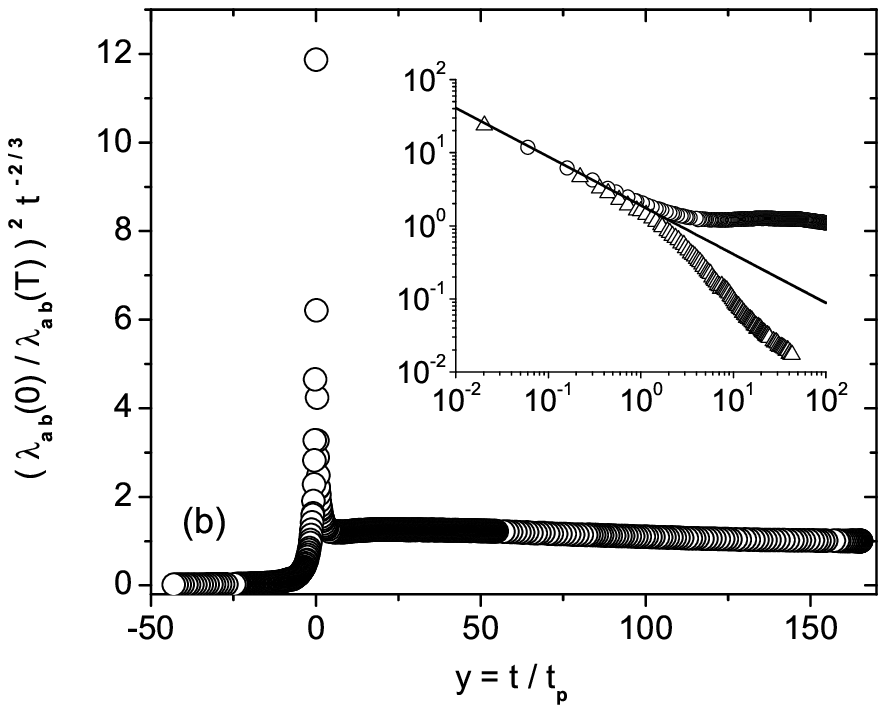}
\caption{(a)Microwave surface impedance data for $\lambda
_{ab}^{2}\left(0\right) /\lambda _{ab}^{2}\left( T\right) $
($\bigcirc $)\ and $d\left( \lambda _{ab}^{2}\left(0\right)
/\lambda _{ab}^{2}\left( T\right) \right) /dT$ ($\square $) versus
$T$ of a high-quality Bi$_{2}$Sr$_{2}$CaCu$_{2}$O$_{8+\delta }$
single crystal taken from Jacobs {\em et al.}
\protect\cite{jacobs}. The solid line is
$\lambda_{ab}^{2}\left(0\right) /\lambda _{ab}^{2}\left( T\right)
=1.2\left( 1-T/T_{c}\right)^{2/3}$ and the dash-dot line its
derivative with $T_{c}=87.5$K, indicating the leading critical
behavior of the homogeneous system. The dotted line is the tangent
to the inflection point at $T_{p}\approx 87$K, where $d\left(
\lambda _{ab}^{2}\left( T=0\right) /\lambda _{ab}^{2}\left(
T\right) \right) /dT$ is maximum; (b) Finite size scaling function
$g\left( y\right) =\left( \lambda _{0ab}/\lambda _{ab}\left(
T\right) \right) ^{2}\left| t\right| ^{-\nu }$ versus
$y=t/\left|t_{p}\right| $ for the data shown in Fig.\ref{fig5}a.
The solid line in the inset is Eq.(\ref{eq14}) with $g_{0}=1.6$.}
\label{fig5}
\end{figure}

Considering again the 3D-XY critical point, extended to the
anisotropic case, the penetration depths and transverse
correlation lengths in directions $i$ and $j$ are universally
related by\cite{book,hohenberg}
\begin{equation}
\frac{1}{\lambda _{i}\left( T\right) \lambda _{j}\left( T\right)
}=\frac{16\pi ^{3}k_{B}T}{\Phi _{0}^{2}\sqrt{\xi _{i}^{t}\left(
T\right) \xi _{j}^{t}\left( T\right) }}.  \label{eq11}
\end{equation}
When the superconductor is inhomogeneous, consisting of
superconducting grains with length scales $L_{i}$, embedded in a
non-superconducting medium,\ $\xi _{i}^{t}$ does not diverge but
is bounded by $\xi _{i}^{t}\xi _{j}^{t}=\left( \xi _{k}^{-}\right)
^{2}\leq L_{k}^{2}$, where $i\neq j\neq k $. The resulting finite
size effect is clearly seen in the microwave surface impedance
data for $\lambda _{ab}^{2}\left( T=0\right) /\lambda
_{ab}^{2}\left( T\right) $ versus $T$ of Jacobs {\em et
al.}\cite{jacobs},\ displayed in Fig.\ref{fig5}a. The solid curve
indicates the leading critical behavior of the homogeneous system.
A characteristic feature of this finite size effect is the
occurrence of an inflection point at $T_{p}\approx 87$K, giving
rise to an extremum in $d\left( \lambda _{ab}^{2}\left( T=0\right)
/\lambda _{ab}^{2}\left( T\right) \right) /dT$. Here
Eq.(\ref{eq11}) reduces to
\begin{equation}
\frac{1}{\lambda _{ab}^{2}\left( T_{p}\right) }\approx
\frac{1}{\lambda _{a}\left( T_{p}\right) \lambda _{b}\left(
T_{p}\right) }=\frac{16\pi ^{3}k_{B}T_{p}}{\Phi _{0}^{2}L_{c}}.
\label{eq12}
\end{equation}
With $\lambda _{ab}\left( T=0\right) =1800$\AA\ as obtained from
$\mu $SR measurements \cite{leem}, $T_{p}\approx 87$K and $\lambda
_{ab}^{2}\left( T=0\right) /\lambda _{ab}^{2}\left( T_{P}\right)
=0.066$ we find $L_{c}\approx 68$\AA , consistent with lower
bounds (\ref{eq10b}) derived from the specific heat data. Thus,
although the superconducting grains are of nanoscale only, a very
small correlation volume makes the critical regime attainable.
Clear evidence for 3D-XY critical behavior was also observed in
epitaxially grown Bi2212 films\cite{osborn}.

To substantiate the occurrence of a finite size effect we explore
the consistency with the associated scaling function. In the
present case it is defined in terms of
\begin{equation}
\left( \frac{\lambda _{0ab}}{\lambda _{ab}\left( T\right) }\right)
^{2}\left| t\right| ^{-\nu }=g\left( y\right) ,\ \ y=sign\left(
t\right) \left| t\right| \left( \frac{L_{c}}{\xi _{0c}^{-}}\right)
^{1/\nu }=sign\left( t\right) \frac{\left| t\right| }{t_{p}}.
\label{eq13}
\end{equation}
For $t$ small and $L_{c}\rightarrow \infty $, so that  $\pm
y\rightarrow \infty $ it should tend to $g\left( y\rightarrow
\infty \right) =1$and $\ g\left( y\rightarrow -\infty \right) =0$,
respectively, while for $t=0$ and $L\neq 0$ it diverges as
\begin{equation}
g\left( y\rightarrow 0\right) =g_{0}y^{-\nu }=g_{0}\left(
\frac{\left| t\right| }{t_{p}}\right) ^{-\nu },  \label{eq14}
\end{equation}
so that in this limit $\left( \lambda _{0ab}/\lambda _{ab}\left(
T_{c},L\right) \right) ^{2}=g_{0}\left| t_{p}\right| ^{\nu
}=g_{0}\xi _{0c}^{-}/L_{c}$. As expected, a sharp superconductor
to normal state transition requires domains of infinite extent.
Moreover at $t_{p}$, $y=1$ and $d\left( \lambda _{0ab}/\lambda
_{ab}\left( T,L\right) \right) ^{2}/dt=0$. Accordingly, there is
an inflection point at $t_{p}$. Since the scaling function
$g\left( y\right) $ depends on the type of confining geometry and
on the conditions imposed (or not, in the case of free boundaries)
at the boundaries of the domains, this applies to the amplitude
$g_{0}$ as well. In Fig.\ref{fig5}b we displayed the finite size
scaling resulting from the data shown in Fig.\ref{fig5}a. In view
of the fact that the data satisfies the limiting behavior of the
finite size scaling function remarkably well, there is strong
evidence for a finite size effect.

\section{Finite size and isotope effects}

Recently we explored the effect of oxygen isotope exchange in
Y$_{1-x}$Pr$_{x}$Ba$_{2}$Cu$_{3}$O$_{7-\delta }$ on the
inhomogeneity induced finite size effect by means of in-plane
penetration depth measurements\cite{tsrkhk}. Defining the relative
oxygen isotope shift of a physical quantity $X$ as $\Delta
X/X=(^{18}X-^{16}X)/^{16}X$) we note that the shifts are not
independent but according to Eq.(\ref{eq12}) related by
\begin{equation}
\frac{\Delta L_{c}}{L_{c}}=\frac{\Delta
T_{p_{c}}}{T_{p_{c}}}+\frac{\Delta \lambda _{ab}^{2}\left(
T_{p_{c}}\right) }{\lambda _{ab}^{2}\left( T_{p_{c}}\right) }.
\label{eq15}
\end{equation}
From the resulting estimates, listed in Table II The resulting
estimates are summarized several observations emerge. First,
$L_{c}$ increases systematically with reduced $T_{p_{c}}$. Second,
$L_{c}$ grows with increasing $x$ and upon isotope exchange
($^{16}$O, $^{18}$O). Third, the relative shift of $T_{p_{c}}$ is
very small. This reflects the fact that the change of $L_{c}$ is
essentially due to the superfluid, probed in terms of $\lambda
_{ab}^{2}$. Accordingly, $\Delta L_{c}/L_{c}\approx \Delta \lambda
_{ab}^{2}/\lambda _{ab}^{2}$ for $x=0,\ 0.2$ and $0.3$. Indeed the
relative shifts of $T_{p_{c}}$, $\lambda _{ab}^{2}\left(
T_{p_{c}}\right) $ and $L_{c} $ are not independent.

\bigskip

\begin{center}
\begin{tabular}{|c|c|c|c|}
\hline x & 0 & 0.2 & 0.3 \\ \hline $\Delta T_{p_{c}}/T_{p_{c}}$ &
-0.000(2) & -0.015(3) & -0.021(5) \\ \hline $\Delta
L_{p_{c}}/L_{p_{c}}$ & 0.12(5) & 0.13(6) & 0.16(5) \\ \hline
$\Delta \lambda _{ab}^{2}\left( T_{p_{c}}\right) /
\lambda_{ab}^{2}\left( T_{p_{c}}\right) $ & 0.11(5) & 0.15(6) &
0.15(5) \\ \hline $^{16}\lambda _{ab}^{2}\left(
^{16}T_{p_{c}}\right) /^{16}\lambda _{ab}^{2}\left( 0\right) $ &
4.4(2) & 4.0(2) & 4.4(2) \\ \hline $^{18}\lambda _{ab}^{2}\left(
^{18}T_{p_{c}}\right) /^{16}\lambda _{ab}^{2}\left( 0\right) $ &
4.9(2) & 4.6(2) & 5.2(2) \\ \hline $^{16}T_{p_{c}}(K)$ & 89.0(1) &
67.0(1) & 52.1(2) \\ \hline $^{18}T_{p_{c}}\left( K\right) $ &
89.0(1) & 66.0(2) & 51.0(2) \\ \hline
$^{16}L_{p_{c}}\left( {\rm \AA} \right) $ & 9.7(4) & 14.2(7) & 19.5(8) \\
\hline
$^{18}L_{p_{c}}\left( {\rm \AA} \right) $ & 10.9(4) & 16.0(7) & 22.6(9) \\
\hline $^{16}\lambda _{ab}\left( 0\right) \left( {\rm \AA} \right)
$ & 1250(10) & 1820(20) & 2310(30) \\ \hline
\end{tabular}
\end{center}

\bigskip

Table II: Finite size estimates for $^{16}T_{p_{c}}$,
$^{18}T_{p_{c}}$, $\ ^{16}\lambda _{ab}^{2}\left(
^{16}T_{p_{c}}\right) /^{16}\lambda _{ab}^{2}\left( 0\right) $ and
$\ ^{18}\lambda _{ab}^{2}\left( ^{18}T_{p_{c}}\right)
/^{16}\lambda _{ab}^{2}\left( 0\right) $, and the resulting
relative shifts $\Delta T_{p_{c}}/T_{p_{c}}$and $\Delta \lambda
_{ab}^{2}\left( T_{p_{c}}\right) /\lambda _{ab}^{2}\left(
T_{p_{c}}\right) $ for an $^{\text{18}}$O content of 89\%.
$^{16}L_{p_{c}}$, $^{18}L_{p_{c}}$ and $\Delta
L_{p_{c}}/L_{p_{c}}$ follow from Eq.(\ref{eq12}). $^{16}\lambda
_{ab}\left( 0\right) $ are $\mu $SR estimates\cite{khasanov}

\bigskip

To appreciate the implications of these estimates, we note that
for fixed Pr concentration the lattice parameters remain
essentially unaffected\cite {conderla,raffa}. Accordingly, an
electronic mechanism, without coupling to local lattice
distortions, implies $\Delta L_{c}=0$. On the contrary, a
significant change of $L_{p_{c}}$ upon oxygen exchange uncovers
the coupling to local lattice distortions involving the oxygen
lattice degrees of freedom. A glance to Table I shows that the
relative change of the domains along the c-axis upon oxygen
isotope exchange is significant, ranging from $12$ to $16$\%,
while the relative change of the inflection point or the
transition temperature is an order of magnitude smaller. For this
reason the significant relative change of $L_{c}$ at fixed Pr
concentration is accompanied by essentially the same relative
change of $\lambda _{ab}^{2}$, which probes the superfluid. This
uncovers unambiguously the existence and relevance of the coupling
between the superfluid, lattice distortions involving the oxygen
lattice degrees of freedom. Potential candidates are the Cu-O
bond-stretching-type phonons showing temperature dependence, which
parallels that of the superconductive order parameter\cite{chung}.
Independent evidence for the shrinkage of limiting length scales
upon isotope exchange stems from the behavior close to the quantum
superconductor to insulator transition where $T_{c}$
vanishes\cite{tsiso}. Here the cuprates become essentially two
dimensional and correspond to a stack of independent slabs of
thickness $d_{s}$\cite{tseuro,tsphysicab}. It was found that the
relative shift $\Delta d_{s}/d_{s}$ upon isotope exchange adopts a
rather unique value, namely $\Delta d_{s}/d_{s}\approx
0.03$\cite{tsiso}. Although the majority opinion on the mechanism
of superconductivity in the cuprates is that it occurs via a
purely electronic mechanism involving spin excitations, and
lattice degrees of freedom are supposed to be irrelevant, the
relative isotope shift $\Delta L_{c}/^{16}L_{c}\approx \Delta
\lambda _{ab}^{2}/^{16}\lambda _{ab}^{2}\approx 0.15$ uncovers
clearly the existence and relevance of the coupling between the
superfluid, lattice distortions and anharmonic phonons which
involve the oxygen lattice degrees of freedom.

In contrast to YBa$_{2}$Cu$_{3}$O$_{7-\delta }$, (K,Ba)BiO$_{3}$,
MgB$_{2}$, 2H-NbSe$_{2}$ and Nb$_{77}$Zr$_{23}$, where the lower
bounds for the length scale $L$ of the superconducting grains
ranges from $182$ A to $814$ \AA , we have seen that the data for
Bi$_{2}$Sr$_{2}$CaCu$_{2}$O$_{8+\delta }$ single crystals and
Y$_{1-x}$Pr$_{x}$Ba$_{2}$Cu$_{3}$O$_{7-\delta }$ uncovers the
existence of nanoscale inhomogeneities, which are not merely an
artefact of the surface, but a bulk property with spatial extent,
giving rise to finite size effects and with that to a rounded
thermodynamic phase transition which occurs smoothly. While there
are many questions to be answered about the intrinsic or extrinsic
origin of the inhomogeneities, the existence and the nature of a
macroscopic superconducting state, we established that a finite
size scaling analysis yields the spatial extent of the
inhomogeneities. Furthermore, we established that the change of
the spatial extent of the inhomogeneities upon oxygen isotope
exchange uncovers the relevance of lattice degrees of freedom in
mediating superconductivity.
\bigskip
\acknowledgments The author is grateful to R. Khasanov, H. Keller
and K.A. M\"{u}ller for very useful comments and suggestions on
the subject matter.

\end{document}